# DEW: A Fast Level 1 Cache Simulation Approach for Embedded Processors with FIFO Replacement Policy

Mohammad Shihabul Haque    Jorgen Peddersen    Andhi Janapsatya    Sri Parameswaran*

University of New South Wales, Sydney, Australia

{mhaque, jorgenp, andhij, sridevan}@cse.unsw.edu.au

## ABSTRACT

Increasing the speed of cache simulation to obtain hit/miss rates enables performance estimation, cache exploration for embedded systems and energy estimation. Previously, such simulations, particularly exact approaches, have been exclusively for caches which utilize the least recently used (LRU) replacement policy. In this paper, we propose a new, fast and exact cache simulation method for the First In First Out(FIFO) replacement policy. This method, called DEW, is able to simulate multiple level 1 cache configurations (different set sizes, associativities, and block sizes) with FIFO replacement policy. DEW utilizes a binomial tree based representation of cache configurations and a novel searching method to speed up sim-ulation over single cache simulators like Dinero IV. Depending on different cache block sizes and benchmark applications, DEW oper-ates around 8 to 40 times faster than Dinero IV. Dinero IV compares 2.17 to 19.42 times more cache ways than DEW to determine accu-rate miss rates.

## 1. INTRODUCTION

Cache memories have been used to effectively reduce the ever in-creasing speed gap between the main memory and the processor. Uti-lizing data and instruction caches in computing systems improves performance while reducing energy consumption.

A processor based embedded system, where an application or a class of applications is repeatedly executed, can be customized by the adroit selection of a suitable cache. Multiple studies [5, 8, 13, 18] have found that the correct combination of different cache parame-ters, such as the cache size, number of cache sets(set size), associativ-ity, cache block size (also known as cache line size), etc. can reduce the energy consumption and increase the overall system performance significantly. Application specific processor design platforms such as Tensilica's Xtensa [2, 15] allows the cache to be customized for the processor to meet tighter energy, performance and cost constraints. A cache system which is too large will unnecessarily consume power and increase access time, while a cache system too small will thrash, reducing performance.

Due to the erratic nature of caches, there is no known way of ac-curately determining hit and miss rates without simulating an appli-cation's trace of memory requests. To simulate the trace on caches with hundreds of differing cache parameters can take several months and is simply not feasible. Therefore, several studies have endeav-ored to speed up simulation of cache memories. Among the simula-tion methods, some approaches simulate caches with all the possible combinations of cache parameters under consideration extensively to maintain reliability(i.e., exact values of hits and misses). These are called 'Exact Approaches'. One of the widely used exact approach based single processor cache simulation tool is Dinero IV [7], de-signed by Jan Elder and Mark Hill. Dinero IV can simulate only a single combination of cache parameters at a time. Among the exact approaches, some approaches [13, 20] are able to simulate multi-ple combination of cache parameters in a single pass directly over an application trace. These approaches mainly depend upon cache inclusion properties to speed up simulation. However, caches with the FIFO (or round robin) policy do not exhibit inclusion properties. Therefore, there has been no known work which attempts to speed up the simulation of multiple caches which implement the FIFO re-placement policy.

As FIFO replacement is inexpensive to implement in the hardware, FIFO is a popular choice for level 1 cache in the embedded pro-cessors (i.e., Xtensa LX2 processor [3] and Intel XScale processors [1]). Besides that, previous studies [4] have shown that for L1 cache (especially, data cache), both FIFO and LRU have their own advan-tages. Therefore, in our research, we have decided to extend the sim-ulation approach for FIFO replacement policy. In our research, we have analyzed the features of FIFO replacement policy that prevent us from establishing fast simulation properties when all the caches under simulation use the FIFO replacement policy. We have studied the exact simulation methods, especially Janapsatya's method with the proposed enhancements in the CRCB algorithm [13, 20], to de-termine how the inclusion properties benefit simulators. Resulting from our findings, we propose a new simulation strategy "Direct Ex-plorer Wave"(DEW) to speed up simulation of multiple combination of cache parameters with FIFO replacement policy in a single pass directly over an application trace.

The rest of the paper is structured as follows. Section 2 presents related works, Section 3 presents the background of our research, Section 4 describes our DEW simulation approach, Section 5 de-scribes the experimental setup and discusses the results found for mediabench applications; and section 6 concludes the paper.

## 2. RELATED WORK

Cache performance evaluation has been studied extensively for a long time to find the optimal combination of cache parameters for level 1 cache in embedded systems. The methods of cache evaluation can be broadly categorized in two: estimation and simulation depen-dent. Estimation approaches [8, 10, 17, 21] depend on heuristics, are fast to compute, but are limited in their accuracy. Simulation based approaches [7, 12, 13] usually produce error free results of cache hits and misses. However, they take a longer time than estimation approaches to execute.

Several techniques are used to make simulation of application traces faster. One such technique is fractional simulation [12, 16], which al-lows the simulation of a section of the trace, and obtains results at the cost of accuracy. Another technique simulates the trace for a number of cache configurations(different combination of cache parameters) simultaneously, and produces exact results. These concurrent simu-lations use the knowledge of cache behavior between configurations to speed up simulation considerably. For example, if a hit occurs in a cache with four sets, it is guaranteed to be a hit on a cache with eight sets, provided that both of the caches use the Least Recently Used (LRU) replacement policy, and have equal associativity and block size.

Due to the reliability, many methods have been proposed to im-prove the speed of exact, concurrent, simulation based cache evalua-tion approaches. In 1989, Hill et al. in [11] studied the effect of asso-ciativity in caches. They introduced a forest simulation technique to simulate alternate direct mapped caches quickly. Another technique used was the all-associativity methodology, based on the "Stack" algorithm described by Gecsei et al. in [9], for simulating alternate direct mapped caches, fully-associative caches and set associative caches. Hill et al. showed that for alternate direct mapped caches, forest simulation strategy is faster than the all-associativity method-ology. In 1995, Sugumar et al. [19] proposed a binomial tree depen-dent cache simulation methodology to improve methods described in [11]. Sugumar's method had a time complexity of $O((\log_2 (X)) \times A)$ for searching, where $X$ and $A$ are size and associativity of the cache respectively. Time complexity of maintaining the tree was $O((\log_2(X)) \times A)$. Sugumar's method was applicable only for LRU replacement policy. Due to its flexibility, Sugumar's method promoted the use of binomial tree in simulation of multiple cache configurations in a single pass, took as its input an application trace. Researchers have continued the use of binomial tree to speed up sim-ulation though the focus has remained only on LRU replacement pol-icy. In 2004, Li et al. [16] proposed an improvement to Sugumar's methodology by introducing a compression method to reduce sim-ulation time. The authors of [16] stated that their method can be modified to accommodate the FIFO replacement policy; however, no modification plan for the FIFO replacement policy was given.

---

*Advisor

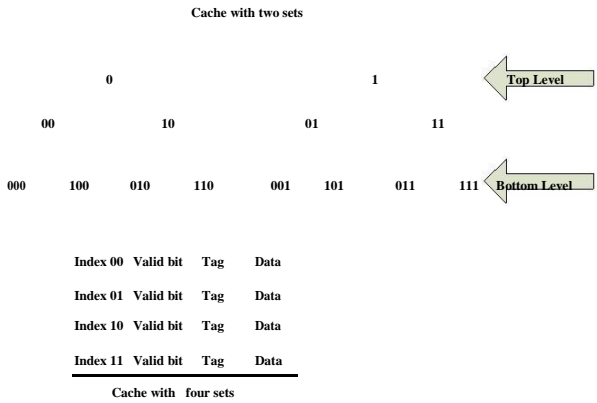

**Figure 1: Formation of simulation tree**

In 2006, Janapsatya et al. [13] proposed a technique by utilizing several LRU based cache inclusion properties and a binomial tree structure. Janapsatya's top-down tree traverse based simulation strat-egy helped to speed up simulation of multiple cache configurations by reading the application trace only once. Janapsatya's searching approach, inside a cache set, took advantage of temporal locality to speed up simulation, as memory address tags were searched accord-ing to their last access time. Therefore, Janapsatya's method had a shorter simulation time than previously proposed solutions. The cache properties and techniques used in Janapsatya's method was exclusive for the LRU replacement policy. Janapsatya's method had a fixed time complexity of $O(\log_2 (X) \times A)$ for searching data or in-structions inside the caches under simulation, where $X$ and $A$ are maximum cache set size and maximum associativity respectively. Time complexity for updating the data structure was $O(\log_2 (X))$. In 2009, Tojo et al. [20] proposed two enhancements to Janapsatya's method in what they called the CRCB algorithm. These pruning based proposals made the simulation even faster by reducing the number of addresses to be examined. The findings of CRCB are also true for FIFO replacement policy; however, the simulation technique was exclusively proposed for the LRU replacement policy.

## 2.1 Contributions and limitations

1. In this paper, we have presented a new simulation strategy "DEW" to simulate multiple level 1 cache configurations of varying set sizes with the FIFO replacement policy by passing over an application trace only once.

2. A novel data structure based on binomial trees and utilizing "wave pointers" has been proposed to enable fast simulation.

3. A search methodology for the above data structure has been proposed, which eliminates unnecessary tag comparisons.

The limitation of DEW is that it is optimized only for the simu-lation of the FIFO replacement policy. It can simulate caches with the LRU replacement policy, but will typically be slower than Janap-satya's method [13] and the CRCB algorithm [20], which are opti-mized only for the LRU policy.

## 3. CACHE PARAMETERS EXPLORATION METHODOLOGY

Cache configurations are mainly parameterized using cache set size ($S$), associativity ($A$) and cache block size ($B$). Cache size ($T$) is the total number of bits that can be stored in the cache. Cache set size ($S$) is the total number of sets in a set associative cache. The number of ways to place data inside a set of a set associative cache is called the associativity($A$). Cache block size($B$), also known as cache line size, is the minimum amount of data that can be stored in a cache. Therefore, $T = S \times B \times A$.

In DEW, we perform simulation on the cache parameters to esti-mate the number of cache misses that would occur for a given col-lection of cache configurations. In DEW, we optimize the run time of simulation by replacing multiple readings of large program traces with a single reading, simulating multiple cache configurations si-multaneously and reducing search complexity inside a cache config-uration. This is possible due to the data structure we have used and the decisions we can make depending on the data structure. In the following subsections, we are going to discuss the data structure used in DEW and the properties that can be used due to the special data structure.

**Figure 3: An address request simulation flow diagram for DEW**

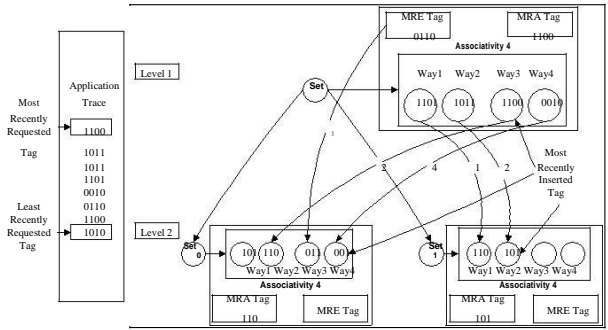

**Figure 4: Simulation tree of DEW after new tag insertion**

## 5. EXPERIMENTAL PROCEDURE AND RESULTS

With the implementation described above, DEW can reduce total simulation time compared to the state of the art cache simulation tool Dinero IV [7]. We have implemented DEW using C++. We have compiled and simulated programs from Mediabench [14] with "SimpleScalar/PISA 3.0d" [6]. Program traces were generated by SimpleScalar and fed into both Dinero IV and DEW. We have verified hit and miss rates of DEW by comparing with Dinero IV and found that they are exactly the same. Simulations were performed on a machine with dual core Opteron64 2GHz processor and 8GB of main memory.

In our implementation of DEW, each tag list is an array and each entry is used to hold a tag (32 bits) and integer wave pointer (32 bits). In total, each tag list entry needs to store 64 bits. In the simulation tree, each node stores the MRA tag (32 bits), MRE tag (32 bits) and wave pointer for the MRE tag (32 bits). Therefore, per tree node or cache set, $(96 + (64 \times A))$ bits are needed, where $A$ is associativity. Thus, per tree level or cache configuration, $(S \times (96 + (64 \times A)))$ bits are needed, where $S$ is the number of sets.

Table 1 shows how the 525 configurations we calculated data for were derived.

| | |
|---|---|
| Cache Set Size=$2^i$ | where $0 <= i <= 14$ |
| Cache Block Size=$2^i$ Bytes | where $0 <= i <= 6$ |
| Associativity=$2^i$ | where $0 <= i <= 4$ |

**Table 1: Cache configuration parameters**

We have simulated cache sizes from 1 byte to 16MB, some of which may be impractical in embedded systems, to have only one tree per forest, and to follow the same experimental methodology

**Algorithm 1** Function Handle_hit()

---
1: N= *position of the cache way which holds the requested tag*;
2: *MRA tag of the current cache set=Requested tag*;
3: *Parent node's Matching entry's wave pointer=N*;
4: *Matching entry location=N*;

---

**Algorithm 2** Function Handle_miss()

---
1: *MRA tag of the current cache set=Requested tag*;
2: *Increase miss counter for the current cache configuration*;
3: N= *position of the cache way which holds the least recently inserted tag*;
4: **if** *The MRA tag of the current cache set is the requested tag* **then**
5:     *Exchange current cache set's $N^{TH}$ cache way's tag and wave pointer with the tag and wave pointer of the MRE entry*;
6: **else**
7:     *Replace current cache set's $N^{TH}$ cache way's tag and wave pointer with the requested tag and "empty"*;
8:     *Update the MRE tag of the current cache set and its wave pointer with the newly evicted tag and its wave pointer*;
9: **end if**
10: *Parent node's Matching entry's wave pointer=N*;
11: *Matching entry location=N*;

---

used in CRCB algorithms [20].

Six Mediabench applications were used to verify the simulators. These are: JPEG encode, JPEG decode, G721 encode, G721 decode, MPEG2 encode and MPEG2 decode. The numbers of memory ad-dress requests have been presented in Table 2 for each of the used applications. All these requests are for byte addressable memory.

| Application | Number of requests |
|---|---|
| Jpeg encode(CJPEG) | 25,680,911 |
| Jpeg decode(DJPEG) | 7,617,458 |
| G721 encode(G721_Enc) | 154,999,563 |
| G721 decode(G721_Dec) | 154,856,346 |
| Mpeg2 encode(MPEG2_Enc) | 3,738,851,450 |
| Mpeg2 decode(MPEG2_Dec) | 1,411,434,040 |

**Table 2: Trace files used for simulation**

Table 3 presents results comparing the DEW simulation approach to Dinero IV [1]. Column 1 lists the applications being simulated. Col-umn 2 shows block size. Columns 3 to 8 show simulation time and columns 9 to 14 show the number of tag comparisons performed by DEW and Dinero IV for different cache associativity. E.g., columns 3 and 4 provide simulation time for DEW and Dinero IV respec-tively to simulate direct mapped cache (1-way) and 4-way set associativity. Direct mapped cache results are used in both cases as DEW auto-matically simulates it while simulating any other associativity. Note that DEW is always much faster than Dinero IV in every case. On average, DEW operates 18 times as fast as Dinero IV. This is due to the significant reduction in tag comparisons.

Figure 5 shows speedup of DEW over Dinero IV based on simu-lation time. Speedup is calculated as the ratio of simulation times. It shows that DEW can run up to 40 times faster than Dinero IV(recorded for JPEG decode, associativity 8 and block size 64 bytes). In the worst case, DEW's run time is still 9 times faster than Dinero IV which was recorded for MPEG2 decode, associativity 4 and cache block size 4 bytes.

Figure 6 shows the percentage reduction of the total number of tag comparisons of DEW over Dinero IV. From Figure 6, it can be seen that Dew can reduce the total number of tag comparisons by 54.9% to 94.9% compared to Dinero IV. DEW reduces 92.97% tag comparisons compared to Dinero IV for JPEG Decode, block size of 64 byte and associativity 4; however, when block size is 4 byte, DEW reduces 70.19% tag comparisons. From Figure 5, it can be seen that speed up of DEW over Dinero IV for these two cases are 39 times and 23 times respectively. The correlation of Figure 5 and Figure 6 illustrates that reduction of tag comparisons helps DEW to reduce total simulation time.

It should be noted that Dinero IV collects different types of in-formation about a cache, such as the number of compulsory misses, number of demand fetches, etc, in addition to cache hit and miss rates. As Dinero IV can simulate only one configuration at a time, to simulate each cache configuration, Dinero IV needs to build the storage for the tags and other information. Maintaining the large in-formation set increases the total simulation time for Dinero IV.

---
[1] Due to space limitations, only limited results are presented.

Table 4 shows the effectiveness of each optimization property used in DEW compared to individual simulation of each cache configura-tion in a simulation forest of DEW without any of the properties de-scribed in Section 3.2 [1]. In this table all the results are for cache with block size of 4 bytes. Column 1 lists the applications being simu-lated. Column 2 shows number of tree nodes needed to be evaluated when only Property 1 (i.e. Binomial tree representation) is used in DEW. This is the worst case number of evaluations for any algorithm. Column 3 shows the total number of simulation tree nodes actually evaluated in DEW using all the four properties of Section 3.2. Col-umn 4 shows how many of the evaluations of Column 3 found the tag in the MRA entry (Property 2); hence avoiding further evaluation of larger set sizes. These three results are associativity independent. Column 5 to 7 and 8 to 10 show, for 4-way and 8-way associativ-ity(including 1-way) respectively, how many times a tag list of a cache set is searched for a requested tag as well as the number of times DEW's properties that avoid searches occurred. For example, column 5 shows total number of tag list searches performed in DEW for associativity 4. Column 6 and 7 show the number of situations, for associativity 4, when a tag list searching is avoided due to hit or miss determined by wave pointer (Property 3) or MRE entry (Property 4) respectively.

From Table 4, it can be seen that the number of node evaluations and the number of situations when a cache set is searched are signifi-cantly smaller when all the properties of DEW are used. The first line of Table 4 can be interpreted as follows. For the JPEG Encode ap-plication, without any optimization, the number of node evaluations would be 770.43 million. However, DEW reduced the total evalua-tions performed to only 140.66 million. This large reduction is due to the property 2 (MRA), which occurred 23.18 million times. Among these 140.66 million evaluations, cache set searching has been per-formed only in 83 million cases for associativity of 4. The reductions arise from the use of properties 3 (Wave) and 4 (MRE) 25.47 million times and 10.24 million times respectively. Therefore, it is evident that the DEW properties are effectively helping to reduce simulation time significantly.

When a tag is available in all the cache configurations in a simu-lation forest, time complexity for DEW's simulation is $O(\log_2(X))$, where X is the maximum cache set size in the search space. If the tag was requested in the previous step, DEW needs only one test. For compulsory misses, time complexity for DEW's simulation is $O(\log_2(X) \times A)$ at best. Dinero IV's time complexity for simula-tion of a tag is $O(\log_2(X) \times A)$ for all the cases.

Therefore, considering all the results and complexities, we can say that DEW shows the fastest performance compared to any other method proposed so far for simulation of level 1 cache with the FIFO replacement policy.

## 6. CONCLUSION

In this paper, we have presented a fast cache simulator, DEW, that can simulate multiple level 1 cache configurations with FIFO replacement policy in a single pass directly over an application trace. Utilizing the features of a binomial tree representation of cache con-figurations, DEW is able to reduce the total number of comparisons by up to 94.9% compared to Dinero IV. As a result, DEW can be almost 40 times faster than Dinero IV. Even in the worst case, DEW is almost 8 times faster than Dinero IV.

This research was supported under Australian Research Council's Discovery Projects funding scheme (Project Number DP0986091).

| Application | Block Size (Bytes) | Total Simulation Time (seconds) | | | | | | No. of tag comparisons (millions) | | | | | |
|---|---|---|---|---|---|---|---|---|---|---|---|---|---|
| | | Assoc 1 & 4 | | Assoc 1 & 8 | | Assoc 1 & 16 | | Assoc 1 & 4 | | Assoc 1 & 8 | | Assoc 1 & 16 | |
| | | DEW | Din. IV | DEW | Din. IV | DEW | Din. IV | DEW | Din. IV | DEW | Din. IV | DEW | Din. IV |
| JPEG enc. | 4 | 30 | 350 | 30 | 357 | 31 | 355 | 357 | 1,397 | 523 | 2,067 | 721 | 3,195 |
| G721 enc. | 4 | 191 | 1,993 | 197 | 2,040 | 220 | 2,036 | 2,656 | 7,921 | 4,382 | 11,401 | 7,170 | 17,152 |
| MPEG2 enc. | 4 | 5,558 | 50,385 | 5,730 | 51,918 | 6,085 | 51732 | 81,691 | 216,232 | 133,165 | 330,678 | 210,704 | 531,065 |
| JPEG dec. | 4 | 10 | 227 | 10 | 229 | 10 | 228 | 122 | 411 | 193 | 599 | 278 | 931 |
| G721 dec. | 4 | 198 | 2,008 | 201 | 2,054 | 225 | 2,052 | 2,710 | 7,942 | 4,406 | 11,393 | 7,289 | 17,235 |
| MPEG2 dec. | 4 | 2,141 | 19,151 | 2,201 | 19,720 | 2,440 | 19,603 | 32,509 | 78,857 | 52,553 | 116,519 | 82,341 | 179,448 |
| JPEG enc. | 16 | 21 | 342 | 22 | 348 | 22 | 349 | 148 | 1,255 | 198 | 1,766 | 280 | 2,649 |
| G721 enc. | 16 | 125 | 1,940 | 127 | 1,972 | 135 | 1,970 | 1,062 | 7,007 | 1,692 | 9,444 | 2,585 | 13,186 |
| MPEG2 enc. | 16 | 3,518 | 48,947 | 3,619 | 50,275 | 3,534 | 50,207 | 31,092 | 192,193 | 47,924 | 275,494 | 70,256 | 419,894 |
| JPEG dec. | 16 | 7 | 221 | 7 | 223 | 7 | 223 | 53 | 364 | 75 | 500 | 101 | 749 |
| G721 dec. | 16 | 132 | 1,954 | 134 | 1,993 | 141 | 1,989 | 1,094 | 7,028 | 1,699 | 9,431 | 2,655 | 13,341 |
| MPEG2 dec. | 16 | 1,337 | 18,479 | 1,350 | 18,958 | 1,429 | 18,914 | 13,264 | 68,287 | 19,932 | 94,703 | 28,500 | 136,879 |
| JPEG enc. | 64 | 19 | 336 | 18 | 342 | 18 | 344 | 76 | 1,161 | 101 | 1,583 | 146 | 2,218 |
| G721 enc. | 64 | 99 | 1,909 | 99 | 1,930 | 101 | 1,932 | 328 | 6,364 | 482 | 8,222 | 692 | 11032 |
| MPEG2 enc. | 64 | 2,732 | 47,813 | 2,729 | 49,076 | 2,488 | 49,325 | 10,893 | 176,249 | 15,184 | 240,811 | 19,953 | 344,404 |
| JPEG dec. | 64 | 6 | 219 | 6 | 220 | 6 | 220 | 23 | 332 | 32 | 437 | 43 | 608 |
| G721 dec. | 64 | 101 | 1,924 | 100 | 1,948 | 105 | 1,960 | 401 | 6,405 | 587 | 8,025 | 821 | 10,614 |
| MPEG2 dec. | 64 | 989 | 18,132 | 983 | 18,480 | 1,018 | 18,564 | 4,837 | 61,783 | 6,700 | 81,505 | 8,156 | 113,118 |

**Table 3: Comparison between Dinero IV and DEW showing simulation time and total number of tag comparisons**

| Application | Unoptimized evaluations | DEW node evaluations | MRA count (Property 2) | Associativity 1 & 4 | | | Associativity 1 & 8 | | |
|---|---|---|---|---|---|---|---|---|---|
| | | | | Searches | Wave count (Property 3) | MRE count (Property 4) | Searches | Wave count (Property 3) | MRE count (Property 4) |
| JPEG enc. | 770.43 | 140.66 | 23.18 | 83.00 | 25.47 | 10.24 | 66.11 | 42.79 | 9.45 |
| JPEG dec. | 228.52 | 46.92 | 7.31 | 28.46 | 8.62 | 2.87 | 24.44 | 14.50 | 0.90 |
| G721 enc. | 4,649.99 | 975.85 | 140.30 | 623.12 | 165.45 | 49.53 | 555.52 | 263.00 | 18.05 |
| G721 dec. | 4,645.69 | 998.35 | 141.07 | 636.09 | 179.16 | 44.51 | 556.95 | 280.05 | 21.09 |
| MPEG2 enc. | 112,165.54 | 28,875.48 | 3,582.20 | 19,213.83 | 4,851.68 | 1,330.80 | 16,635.70 | 8,122.43 | 591.16 |
| MPEG2 dec. | 42,343.02 | 11,465.94 | 1,394.73 | 7,640.57 | 1,964.88 | 507.92 | 6,552.25 | 3,333.98 | 212.69 |

**Table 4: Effectiveness of properties used in DEW (all results in millions)**

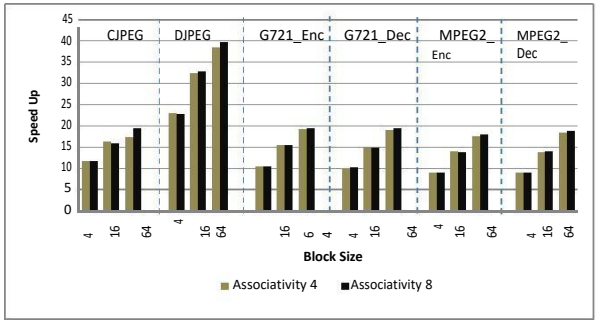

**Figure 5: Speed up of DEW over Dinero IV**

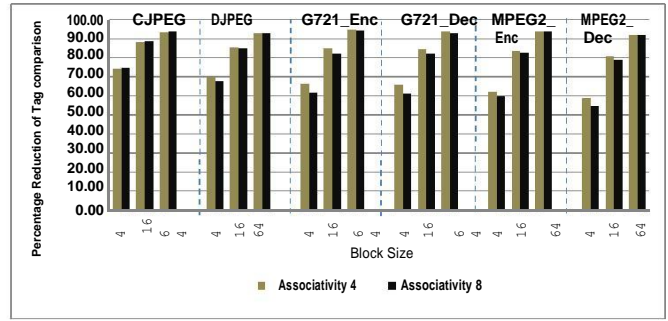

**Figure 6: Reduction of tag comparison in DEW**